\documentclass[prb,showpacs,twocolumn,floatfix,bibnotes]{revtex4}

\usepackage{here}
\usepackage{graphicx}
\usepackage{bm}

\newcommand{\be}{\begin{equation}}
\newcommand{\ee}{\end{equation}}

\newcounter{Fig}

\begin{document}
\begin{sloppy}

\title{Nonlinear surface impurity in a semi-infinite 2D square lattice}

\author{Mario I. Molina}

\affiliation{Departamento de F\'{\i}sica, Facultad de Ciencias,
Universidad de Chile, Casilla 653, Santiago, Chile} 

\begin{abstract}
We examine the formation of localized states on a generalized
nonlinear impurity located at, or near the surface of a
semi-infinite 2D square lattice. Using the formalism of lattice Green
functions, we obtain in closed form the number of bound states as
well as their energies and probability profiles, for different
nonlinearity parameter values and nonlinearity exponents, at
different distances from the surface. We specialize to two cases:
impurity close to an ``edge'' and impurity close to a ``corner''.
We find that, unlike the case of a 1D semi-infinite lattice, in 2D, 
the presence of the surface helps the formation of a localized state.
\end{abstract}


\maketitle

\noindent

An interesting recent development for extended, nonlinear systems
with discrete translational invariance is the the concept of
``breather'' or ``intrinsic localized mode'', whose existence is
the result of a careful balance between nonlinearity and
discreteness\cite{kv_pt}. These excitations are thought of as
generic to a wide range of different physical systems, including
Josephson junctions\cite{junction}, biopolymers\cite{bio},
Bose-Einstein condensates in a magneto-optical trap\cite{BE} and
arrays of nonlinear optical waveguides\cite{DS}, among others. In
nonlinear optics, these excitations are known as ``discrete
solitons''(DS) due to their ability to move in a more or less
robust manner, when endowed with momentum (beam angle). In fact,
many theoretical predictions made for DS have now been
experimentally verified in optics, causing a surge of activity in
this field. It is believed that an understanding on the creation
and propagation of DS under different conditions, might have a
substantial impact on future telecomunication/computing systems.

When looking for discrete solitons, one notes that in the limit of
high nonlinearity or high power, the effective nonlinearity is
concentrated in a few ``sites'' only and, therefore, it makes
sense to make the approximation of replacing the whole nonlinear
system for a simpler one, consisting of a discrete linear lattice
with a a small nonlinear cluster, or even a single site embedded
in it. The simplified system is oftentimes amenable to exact
mathematical treatment, leading to closed-form expressions for the
relevant energies and nonlinearity parameters, as well as
providing a bound state spatial profile for the relevant
amplitudes, be these electronic or optical. This high-nonlinearity
localized state provides a very good starting point when looking
for discrete solitons in a more general, less restrictive
context.\cite{prb1,prb2}.

On the other hand, given the practical need to scale down the
components of any all-optical system, such as waveguide arrays, it
becomes important to understand how the presence of some realistic
effects such as boundaries or surfaces affect the creation and
propagation characteristics of these DS. Discrete surface solitons
at the edge of a one-dimensional (1D) waveguide array has been
predicted\cite{makris} and experimentally observed\cite{suntsov}.
It has been shown that the presence of nonlinearity can stabilize
the surface modes in discrete systems, and give rise to different
types of states localized at or near a 1D surface, in a
vibrational\cite{takeno} or optical context\cite{mvk_OL}.
\begin{figure}[t]
\noindent\includegraphics[scale=0.375]{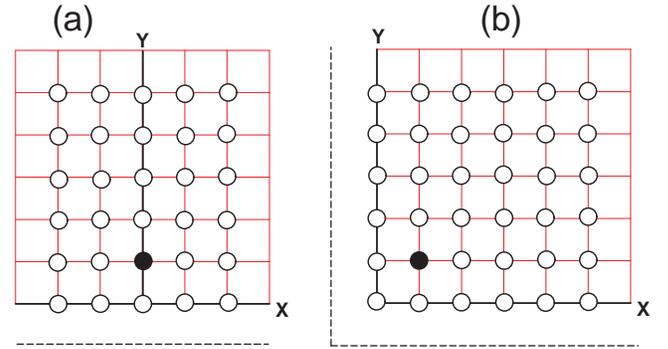}
\caption{(color online) Nonlinear impurity placed near the
``side'' (a) and near the ``corner'' (b) of a semi-infinite square
lattice. Along the dashed lines the amplitude is strictly zero.}
\label{fig1}
\end{figure}

In this work, we consider surface effects for a  simple
two-dimensional (2D) system consisting of a nonlinear impurity
placed near the boundary of a semi-infinite square lattice(Fig.1),
and examine the conditions for the existence of bound state(s),
and compare them to the results obtained for 1D case.

The stationary modes of a $D$-dimensional discrete lattice in the
presence of a single nonlinear impurity located at ${\bf d}$ are
obtained from the stationary-state discrete nonlinear
Schr\"{o}dinger (DNLS) equation
\be -i \beta C_{\bf n} + V\; \sum_{n.n}\; C_{\bf m} + \delta_{{\bf
n d}}\chi\; | C_{\bf n} |^{\alpha}\; C_{\bf n} = 0 \label{eq:dnls}
\ee
where ${\bf n}$ is a site of a $D$-dimensional lattice, $V$ is the
transfer matrix element, $\chi$ is the nonlinearity parameter and
$\alpha$ is the nonlinearity exponent. The sum in (\ref{eq:dnls})
is usually restricted to nearest-neighbors, but other cases have
also been considered\cite{prb_dispersion}. In the {\em
conventional} DNLS case, $\alpha = 2$ and $\chi$ is proportional
to the square of the electron-phonon coupling at site ${\bf n}$,
while $\beta$ is the eigenenergy. In nonlinear optics
Eq.(\ref{eq:dnls}) describes the transversal dynamics of an
optical field in an array of weakly coupled linear waveguides, in
the presence of a single, nonlinear (Kerr) waveguide. There,
$\alpha = 2$, $C_{n}$ is the normalized amplitude of the field in
the nth waveguide, while V is the coupling among waveguides and
$\chi$ is the effective nonlinearity of the ``impurity'' waveguide
proportional to the nonlinear Kerr coefficient. Also in this case,
$\beta$ in Eq.(\ref{eq:dnls}) must be understood as the
propagation constant for the allowed optical modes along the
longitudinal coordinate of the array. Hereafter, for the sake of
definiteness, we will work in a condensed matter context, but the
results obtained can be applied to nonlinear optics, where
appropriate.

\section{Localized states near the surface of a 2D square lattice}

Let us examine the existence of bound states around a single
generalized nonlinear impurity located near the surface of a
semi-infinite square lattice (Fig.1(a), (b)). We follow in this
section the Green function procedure already described in previous
works\cite{prb1,prb2}, so that the reader already familiar with
this formalism can skip this section and proceed directly to the
next one. We denote by ${\bf d} = (d_{x}, d_{y})$ the position of
the impurity. By normalizing all energies to the half bandwidth of
the infinite chain case ($4 V$), the dimensionless Green function
$G = 1/(z - H)$ can be formally expanded as\cite{economou}
    \be
G = G^{(0)} + G^{(0)} H_{1} G^{(0)} + G^{(0)} H_{1} G^{(0)} H_{1}
G^{(0)} + \cdots\label{eq:G}
    \ee
where $G^{(0)}$ is the unperturbed ($\chi = 0$) Green function of
the semi-infinite lattice and $H_{1} = \gamma |C_{\bf d}|^{\alpha}
|{\bf d}\rangle\langle {\bf d}|$, with $\gamma\equiv \chi/4V$.
Series (\ref{eq:G}) can be resumed to all orders to yield
\be G_{{\bf m} {\bf n}} = G_{{\bf m} {\bf n}}^{(0)} + {\gamma
|C_{{\bf d}}|^{\alpha}G_{{\bf m} {\bf d}}^{(0)} G_{{\bf d} {\bf
n}}^{(0)}\over{1 - \gamma |C_{{\bf d}}|^{\alpha} G_{{\bf d} {\bf
d}}^{(0)}}},\label{eq:Gmn} \ee
where $H_{{\bf m} {\bf n}} \equiv \langle {\bf m}|G|{\bf
n}\rangle$. The energy of the bound state(s) is obtained form the
poles of $G_{{\bf m} {\bf n}}$, i.e., by solving
\be 1 = \gamma |C_{{\bf d}}|^{\alpha} G_{{\bf d} {\bf
d}}^{(0)}(z_{b}), \label{zb} \ee
while the bound state amplitudes $C_{{\bf n}}$ are obtained form
the residues of $G_{{\bf m} {\bf n}}$ at $z = z_{b}$:
\be |C_{{\bf d}}|^{2} = -{G_{{\bf n} {\bf d}}^{(0)}(z_{b}) G_{{\bf
d} {\bf n}}^{(0)}(z_{b})\over{{G'}_{{\bf d} {\bf
d}}^{(0)}(z_{b})}} \label{cb} \ee
Inserting this back into the bound state energy equation leads to
a nonlinear equation for the eigenenergies:
\be {1\over{\gamma}} = {{G_{{\bf d} {\bf d}}^{(0)}}^{\alpha +
1}(z_{b})\over{[-G_{{\bf d} {\bf d}}^{'(0)}
(z_{b})]^{\alpha/2}}}.\label{eq:zb} \ee
The unperturbed Green function $G_{{\bf m} {\bf n}}^{(0)}$ for the
semi-infinite lattice, can be calculated by a judicious
application of the method of mirror images, as we will show in the
next two sections.

\section{Impurity close to an ``edge''}

We start by placing the impurity near the edge of the lattice as
depicted on Fig.1(a). In order to simplify matters, we take ${\bf
d} = (0,d)$. Since there is no lattice below $(0,0)$, $G_{{\bf m}
{\bf n}}^{(0)}$ should vanish identically along the sites lying on
the dashed line in Fig.1(b). This implies,
\be
G_{{\bf d} {\bf d}}^{(0)}  = G_{{\bf d} {\bf d}}^{\infty} -
G_{{\bf d},{-{\bf d}-2{\bf j}}}^{\infty}.
\ee
where ${\bf j}$ is a unit vector in the $y$-direction and where
$G_{{\bf m} {\bf n}}^{\infty}$ refers to the Green function of the
infinite 2D square lattice. Now, using the translation invariance
property $G_{{\bf m}{\bf n}}^{\infty} = G_{{\bf m}-{\bf
n}}^{\infty}$, and the symmetry $G_{{\bf m}{\bf n}}^{\infty} =
G_{{\bf n}{\bf m}}^{\infty}$, we have $G_{{\bf d} {\bf
d}}^{\infty} = G_{{\bf 0} {\bf 0}}^{\infty}$ and $G_{{\bf
d},{-{\bf d}-2{\bf j}}}^{\infty} = G_{{\bf 0},{2 {\bf d} + 2{\bf
j}}}^{\infty}$ or, using a simplified notation,
\be
G_{{\bf d} {\bf d}}^{(0)} =  G(z; 0, 0) - G(z; 0, 2 d +
2)\label{eq:Gside}
\ee
where $G(z;m, n)$ refers to the Green function for an infinite
square lattice
\be
G(z;m,n) =
{1\over{\pi^{2}}}\int_{0}^{\pi}d\phi_{1}\int_{0}^{\pi}d\phi_{2}
{\cos(m \phi_{1})\cos(n \phi_{2})\over{z - (1/2) (\cos(\phi_{1}) +
\cos(\phi_{2}))}}\label{eq:2DG}
\ee
(see, for instance, ref.\cite{morita}). We note that
(\ref{eq:Gside}) is identically zero at $d = -1$. The computation
of $G(z; 0, d)$ and $G(z; 0, 2 d +2)$ can be achieved by using
some recurrence relations\cite{morita} by means of which, an
arbitrary Green function $G(z;m,n)$ can be expressed in terms of
two Green functions only: $G(z;0,0)$ and $G(z;1,1)$, where
$G(z;0,0) = (2/\pi z) K[1/z^{2}]$ and $G(z;1,1) = (2/\pi z) (\
(2z^{2} - 1)K[1/z^{2}]-2 z^{2} E[1/z^{2}]\ )$, where $K[x]$ is the
complete elliptical integral of the first kind:
$K[x]=\int_{0}^{\pi/2} [1 - x \sin(\phi)^{2}]^{-1/2} d\phi$, and
$E[x]$ is the complete elliptical integral of the second kind:
$E[x] = \int_{0}^{\pi/2} [1 - x \sin(\phi)^{2}]^{1/2} d\phi$. In
this way, we have obtained a number of non-diagonal Green
functions in explicit form (see Appendix I). In particular, we
have obtained $G(z; 0,2)$, $G(z; 0,4)$ and $G(z; 0,6)$ and $G(z;
0,6)$ in closed form, needed in Eq.(\ref{eq:Gside}). We finally
insert (\ref{eq:Gside}) into the RHS of the eigenvalue equation,
Eq.(\ref{eq:zb}), and solve for $z_{b}$ numerically. However, the
most important features can be already deduced from the structure
of Eq.(\ref{eq:zb}): In Fig.2 we show
\begin{figure}
\noindent\includegraphics[scale=0.65]{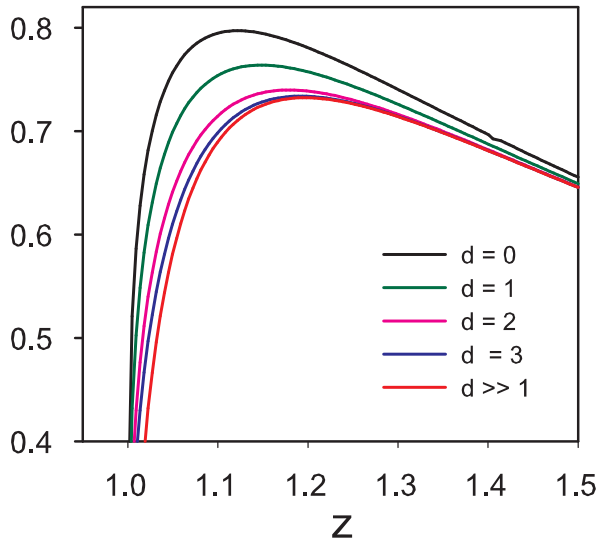} \caption{(color
online) Impurity close to edge: Right-hand side of Eq.(\ref{eq:zb}) versus $z$, for
$\alpha = 2$ and for different distances from the ``edge''.}
\label{fig2}
\end{figure}
the right-hand side of Eq.(\ref{eq:zb}), for the important case
$\alpha = 2$ (standard DNLS) and for different $d$ values. For
comparison, the case $d\rightarrow \infty$ has also been included.
Since it is the intersection of these curves with the horizontal
line $1/\gamma$ what determines the existence of bound states, we
see that in general, for finite $d$ a minimum value of
nonlinearity $\gamma$ is needed to create a bound state. An
increase past the threshold value creates two bound states. One of
these tends to approach the band while the other departs from the
band as $\gamma$ is increased. As argued before in previous
works\cite{prb1,prb2}, the former should correspond to an unstable
localized state, while the latter denotes a stable bound state.

In Fig. 3 we show a bound state phase diagram in nonlinearity
strength-nonlinearity exponent space, showing the number of bound
states, for different positions of the impurity. As the impurity
is brought more and more inside the lattice, the region in
parameter space where two bound states are possible increases. In
the limit $d\rightarrow \infty$, the curve where a single bound
state is found, touches the origin, and coincides with the curve
for an infinite square lattice computed in a previous
work\cite{mm_prb_square}, as expected.

An interesting question now is, how does the critical nonlinearity
to form a localized state $\gamma_{c}$ depend on $d$? Such
critical nonlinearity value is formally given by the inverse of
the RHS of Eq.(\ref{eq:zb}), evaluated at precisely the value of
$z$ where the RHS of Eq.(\ref{eq:zb}) possesses a maximum. In Fig.
4 we show $\gamma_{c}$ versus $d$, for a variety of nonlinearity
exponents. As $d$ is increased past $3$, all curves seem to
converge pretty quickly to their asymptotic values.

The situation depicted in Figs.3 and 4 is qualitatively similar to
what one encounters when placing a nonlinear impurity near the
edge of a semi-infinite 1D lattice\cite{prb1}, with a difference,
though: In the 1D case, for $\alpha = 2$ the presence of the
surface tended to increase $\gamma_{c}$, while in our case, the
proximity of the ``edge'' tends to {\em decrease} $\gamma_{c}$:
its presence helps localization of the excitation. We also observe
that, for a given impurity position $d$, the nonlinearity needed
to create a bound state increases with $\alpha$. This was also
observed in 1D and the explanation is quite general, independent
of dimensionality: From Eq.(\ref{eq:dnls}) we see that, since
$|C_{n}| <1$, as $\alpha$ is increased,$|C_{n}|^{\alpha}$ will
necessarily decrease, meaning that a larger value of $\gamma$ will
be needed to keep the value of the effective impurity strength,
$\gamma |C_{\bf d}|^{\alpha}$.

\begin{figure}
\noindent\includegraphics[scale=0.6]{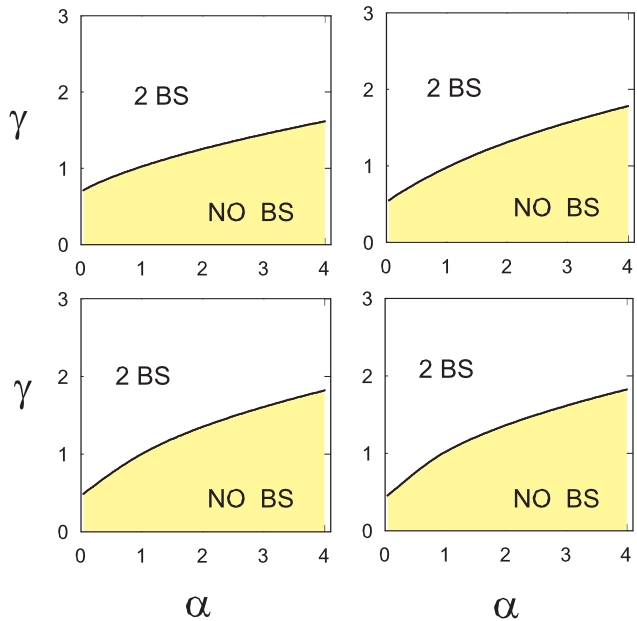} \caption{(color
online) Bound state phase diagrams in nonlinearity
parameter-nonlinearity exponent, for impurity placed at different
distances from the ``edge'': $d=0$(top left), $d=1$ (top right), $d=2$
(bottom left) and $d=3$ (bottom right). On the solid curve, there
is precisely one bound state.} \label{fig3}
\end{figure}

\begin{figure}
\noindent\includegraphics[scale=0.5]{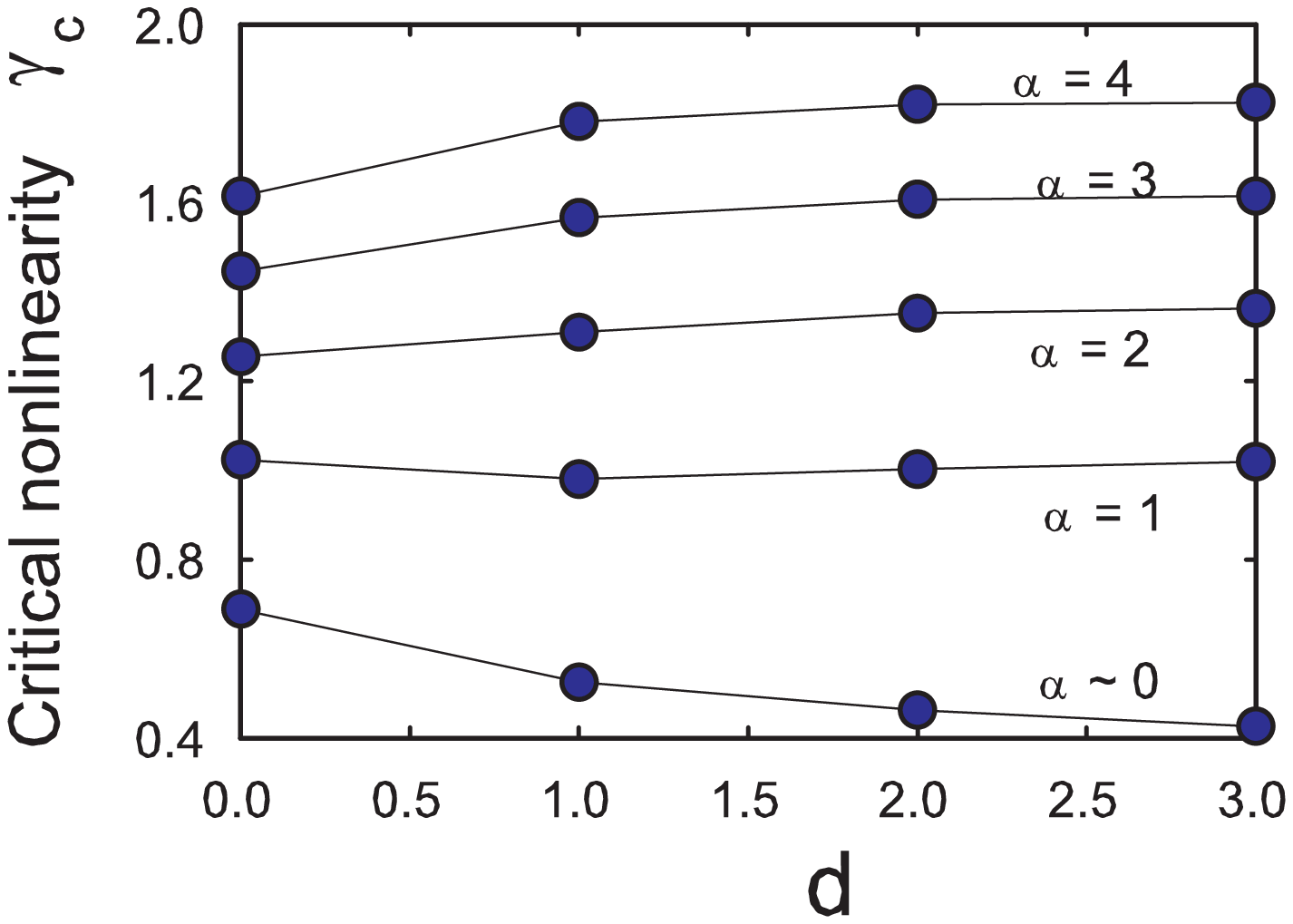} \caption{(color
online) Scaled critical nonlinearity for onset of bound state as a
function of the distance from the nonlinear impurity to the ``edge'' 
of the lattice, for several nonlinearity exponents, ranging from
$\alpha\approx 0$ up to $\alpha=4$.} \label{fig4}
\end{figure}

\section{Impurity close to a ``corner''}
\label{corner}

\begin{figure}[t]
\noindent\includegraphics[scale=0.7]{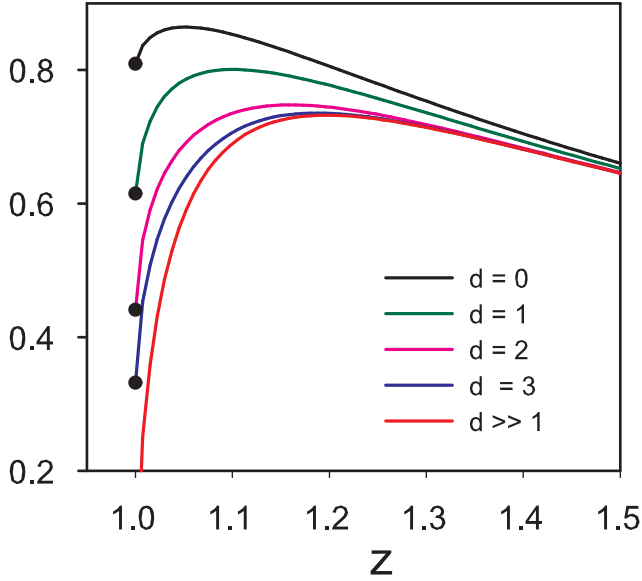} \caption{(color
online) Impurity close to corner: Right-hand side of Eq.(\ref{eq:zb}) versus $z$, for
$\alpha = 2$ and for different distances along the diagonal. }
\label{fig5}
\end{figure}

In this case, the impurity is located near the corner of the
lattice as depicted on Fig.1(b). In order to simplify matters, we
take ${\bf d} = (d,d)$; i.e., we place the impurity along the
``diagonal'' sites. In this case because the impurity is
surrounded by ``more surface'' than in the previous case, one
would expect even stronger departures from the 1D results already
explored in refs.\cite{prb1,prb2}. Since there is no lattice to
the left or below $(0,0)$, $G_{{\bf m} {\bf n}}^{(0)}$ should
vanish identically along the sites lying on the dashed line in
Fig.1(a). Thus,
\begin{eqnarray}
G_{{\bf d},{\bf d}}^{(0)} & = & G_{{\bf d}, {\bf d}}^{\infty} -
G_{{\bf d},(d_{x},-d_{y}-2)}^{\infty} \nonumber\\
                          &   & - G_{{\bf d}, (-d_{x}-2,d_{y})}^{\infty} +
                          G_{{\bf d}, (-d_{x}-2,-d_{y}-2)}^{\infty}\label{eq:Gdd}
\end{eqnarray}
 We can recast
Eq.(\ref{eq:Gdd}) as
\be
G_{{\bf d} {\bf d}}^{(0)}(z) =  G(z;0,0) - 2 G(z;0, 2 d + 2) -
G(z;2 d +2,2 d +2)\label{eq:Gsimplified}
\ee where $G(z;m,n)$ is given by Eq.(\ref{eq:2DG}).
On Fig.5 we show the right-hand side of Eq.(\ref{eq:zb}), for the
important case $\alpha = 2$ (standard DNLS) and for different $d$
values. For comparison, the case $d\rightarrow \infty$ has also
been included. We note an important difference with the case of
the previous section: As $z\rightarrow 1^{+}$, the RHS of
Eq.(\ref{eq:zb}) approaches a finite, non-zero value. This implies
the following: An increase past a minimum value of nonlinearity
$\gamma_{c}^{(1)}$ creates two bound states. One of these tend to
depart from the band while the other approaches the band as
$\gamma$ is increased. The former state is stable while the latter
is unstable and, in fact ceases to exist altogether when $\gamma$
reaches a second critical value $\gamma_{c}^{(2)}$, marked with a
dot on Fig.5. Afterwards, there is only a single bound state. The
value of this second critical nonlinearity can be obtained in
closed form by taking the limit $z\rightarrow 1^{+}$ in
Eq.(\ref{eq:zb})(see Appendix II). In this way,we have obtained:
\begin{eqnarray}
\gamma_{c}^{(2)}(d = 0) & = & 1.236,\hspace{0.5cm}\gamma_{c}^{(2)}(d = 1) = 1.626\nonumber\\
\gamma_{c}^{(2)}(d = 2) & = &
2.267,\hspace{0.5cm}\gamma_{c}^{(2)}(d = 3) = 3.01 \label{eq:gc}
\end{eqnarray}

As $d$ increases, this critical nonlinearity parameter increases
rapidly, and tends to diverge for $d\rightarrow \infty$, that is,
for an infinite square lattice, the unstable bound state will
still be present at arbitrarily large nonlinearity parameter
values, a well-known fact\cite{mm_prb_square}.

In Fig. 6 we show bound state phase diagrams in nonlinearity
strength-nonlinearity exponent space, showing the number of bound
states, for different positions of the impurity. As the impurity
is brought more and more inside the lattice, the region in
parameter space where two bound states are possible increases. In
the limit $d\rightarrow \infty$, the region comprising one bound
state will get more and more ``squeezed'' into the $\gamma$ axis
and will formally disappear for a truly infinite square
lattice\cite{mm_prb_square}.
\begin{figure}[t]
\noindent\includegraphics[scale=0.4]{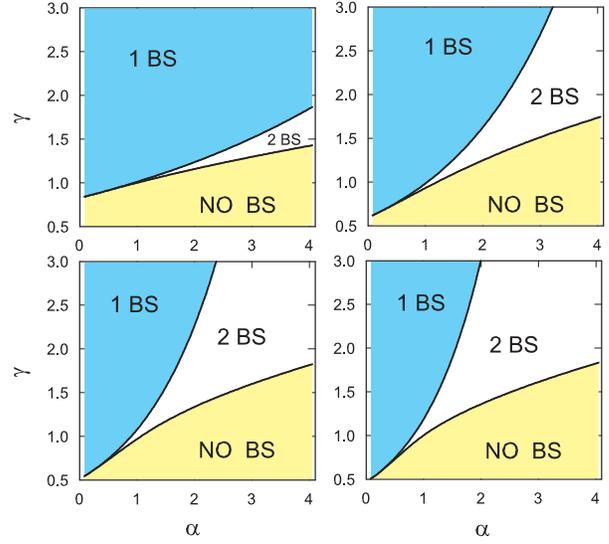} \caption{(color
online) Bound state phase diagrams in nonlinearity
parameter-nonlinearity exponent, for impurity placed at different
(diagonal) distances from the ``corner'': $d=0$(top left), $d=1$ (top
right), $d=2$ (bottom left) and $d=3$ (bottom right). }
\label{fig6}
\end{figure}

 In Fig.7 we show $\gamma_{c}^{(1)}$ versus $d$, for a variety of
nonlinearity exponents. We note that, as the impurity is brought
closer and closer to the ``corner'', $\gamma_{c}^{(1)}$ increases
or decreases, depending on whether $\alpha$ is above or below,
approximately, two. This feature was also present in the previous
case (see Fig.4). However, in this case, the proximity effect of
the corner is much more pronounced. In particular, as the impurity
is brought closer to the corner, the nonlinearity needed to create
a bound state decreases even more than when the impurity is
brought closer to the edge. This implies an even greater departure
from the 1D results. As $d$ is increased past $3$, all curves seem
to start converging towards their asymptotic values.

\begin{figure}
\noindent\includegraphics[scale=0.5]{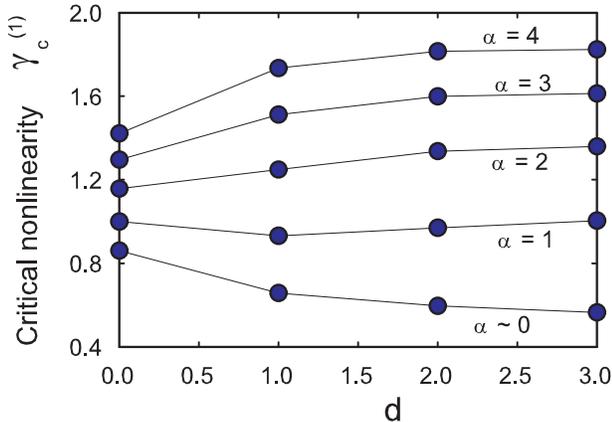} \caption{(color
online) Scaled critical nonlinearity for onset of bound state as a
function of the distance from the nonlinear impurity to ``corner''
of the lattice, for several nonlinearity exponents, ranging from
$\alpha\approx 0$ up to $\alpha=4$.} \label{fig7}
\end{figure}

\section{Conclusion}
We have examined the formation of bound states around a general
nonlinear impurity located at, or near the ``edge'' or the
``corner'' of a semi-infinite 2D square lattice. By means of the
lattice Green functions formalism, we have obtained in closed form
the nonlinear equation for the bound state energies, from which we
have obtained bound state phase diagrams in nonlinearity
strength-nonlinearity parameter, for different impurity positions
with respect to the surface. In general, one finds that, a minimum
value of nonlinearity is needed to create a bound state. Up to two
bound states are possible, although only one of them is always
unstable. These features have been observed previously for the 1D
semi-infinite system. However, for the standard DNLS case
($\alpha = 2$), some interesting departures from the 1D case 
were also found: 
(i) The increased number of surface
sites surrounding the impurity when it is close to the ``corner''
seem to obliterate completely the unstable bound state, for
relatively high nonlinearity values. (ii)  As the impurity is
brought closer to the surface, the nonlinearity needed to create a
localized state {\em decreases}, specially in the case when the
impurity is near a ``corner''. This suggests that, in a more
general context, when considering the creation of discrete
solitons near the surface of a completely nonlinear (Kerr) 2D square
lattice, the surface (edge, corner) of the square lattice would
exert an attractive potential, instead of the repulsive one
observed in semi-infinite 1D systems\cite{mvk_OL}. This would make
the creation of discrete solitons easier to accomplish and observe
near the boundaries of 2D discrete periodic systems.
\vspace{0.5cm}

\section{Acknowledgments}
This work was partially supported by  Fondecyt grants 1050193 and
7050173. The author is grateful to Y. S. Kivshar for useful
discussions.

\section{APPENDIX I}

For an infinite square lattice there are a number of recursion
relations that allow one to express any desired Green function, in
terms of , ultimately two basic ones. We state some recursion
relations (see for instance, Morita\cite{morita}). For the sake of
space, we drop mention of the re-scaled frequency $z$ inside the
argument of the Green functions and define:
\be
A \equiv G(0,0) = {2\over{\pi z}} K[1/z^{2}]
\ee
\be
B = G(1,1) = {2\over{\pi z}} [\  (2 z^{2} - 1) K[1/z^{2}]-2 z^{2}
E[1/z^{2}] \ ]
\ee

and use the relations:
\begin{eqnarray}
G(1,0)      & = & z G(0,0) - 1\\
G(m+1, m+1) & = & {4 m \over{2 m +1}} (2 z^{2} -1) G(m, m) \nonumber\\
            &   &- {2 m -1\over{2 m +1}} G(m-1, m-1)\\
G(m+1,m)    & = & 2 z G(m, m) - G(m, m-1) \ \ \ \ \\
G(m+1,n)    & = & 4 z G(m, n) - G(m-1,n)\nonumber\\
            &   &- G(m, n+1) - G(m, n-1)\\
G(m+1,0)   & = & 4 z G(m,0) - G(m-1,0) \nonumber\\
            &   & - 2 G(m,1)
\end{eqnarray}
Using these relations, one obtains:
\be
G(0,1) = A z -1
\ee
\be
G(0,2) = -2\,B - 4\,z + A\,\left( -1 + 4\,z^2 \right)
\ee
\be
G(1,2) = 1 - A\,z + 2\,B\,z
\ee
\be
G(0,3) = -1 - 12\,B\,z - 16\,z^2 + A\,z\,\left( -3 + 16\,z^2
\right)
\ee
\be
G(2,2) = (1/3)(-A - 4\,B + 8\,B\,z^2)
\ee
\begin{eqnarray}
\lefteqn{G(0,4) =}\\
& &  (1/3) A ( -5 + 192\,z^4 )\nonumber\\
& & -(8/3) ( B + 22\,B\,z^2 + 6\,( z + 4\,z^3 ))
\end{eqnarray}
\be
G(1,3) = (4/3) ( A + 6\,z - 6\,A\,z^2 + B ( (7/4) + 4\,z^2 ) )
\ee
\be
G(2,3) = (1/3) ( -3 + A\,z + 2\,B\,z\,( -7 + 8\,z^2 ) )
\ee
\be
G(3,3) = \frac{-3\,B}{5} + \frac{8\,\left( -1 + 2\,z^2 \right)
\,\left( -A - 4\,B + 8\,B\,z^2 \right) }{15}
\ee
\begin{eqnarray}
G(1,4) & = & 1 + 48\,z^2 + 8\,B\,z\,( 3 + 2\,z^2 )\nonumber\\
       &   & + A ( 9\,z - 48\,z^3 )
\end{eqnarray}
\begin{eqnarray}
G(3,4) & = & 1 + (1/15) A ( 11\,z - 32\,z^3)\nonumber\\
       &   & + (4/15) B\,z\,( 29 - 84\,z^2 + 64\,z^4 )
\end{eqnarray}
\begin{eqnarray}
G(4,4) & = & (64/105) A ( -(71/64) + 6\,z^2 - 6\,z^4 )  \nonumber\\
       &   & +  (8/105) B ( -(11/32) +(59/16)\,z^2 \nonumber\\
       &   &- 9\,z^4 + 6\,z^6 )
\end{eqnarray}
\begin{eqnarray}
G(0,5) & = &(16/3) 48\,z^4\,( -1 + A\,z )  \nonumber\\
       &   & + (1/48)( -3 - 65\,A\,z - 140\,B\,z ) \nonumber\\
       &   & + (1/3) z^2\,( -27 + 15\,A\,z - 50\,B\,z )
\end{eqnarray}
\begin{eqnarray}
G(2,4) & = & (1/15) A\,( -23 + 156\,z^2 )  -(180/15) z \nonumber\\
       &   & + (B/15)( -19 - 136\,z^2 + 96\,z^4)
\end{eqnarray}
\begin{eqnarray}
G(1,5) & = & 8 z ( 3 + 32\,z^2 )  + (1/15) A ( 28 + 504\,z^2 - 3840\,z^4)\nonumber\\
       &   & + (1/15) B ( 43 + 2512\,z^2 + 768\,z^4 )
\end{eqnarray}
\begin{eqnarray}
G(0,6) & = & (1/15) A ( -31 - 2308\,z^2 + 11520\,z^4 +
15360\,z^6)\nonumber\\
       &   &  -(2/15)( 30\,z\,( 9 + 256\,z^2 + 256\,z^4 )\nonumber\\
       &   &  + (1/15) B ( 23 + 3472\,z^2 + 8768\,z^4 )
\end{eqnarray}

\section{APPENDIX II}

For an impurity close to a ``corner'', there is a critical
nonlinearity value $\gamma_{c}^{(2)}$, beyond which, the unstable
bound state ceases to exist. It can be computed by taking the
limit $z_{b}\rightarrow 1^{+}$ in Eq.(\ref{eq:zb}), with the Green
functions obtained in section II. In this way, we have obtained:
\begin{eqnarray}
\gamma_{c}^{(2)}(d = 0) & = & {27\over{64}} \pi^{2} {(4 -
\pi)\over{(3
\pi - 8)^{3}}} = 1.236\nonumber\\
\gamma_{c}^{(2)}(d = 1) & = & {8575 \pi^{2} (65 \pi -
208)\over{1024
(544 - 175 \pi)^{3}}} = 1.626\nonumber\\
\gamma_{c}^{(2)}(d = 2) & = & {132068475 \pi^{2} (647955
\pi-2037676)\over{64(5668760-1805265 \pi)^{3}}} = 2.267\nonumber\\
\gamma_{c}^{(2)}(d = 3) & = & {6087156075 \pi^{2} (74669595 \pi -
234592192)\over{16384 (132029312-42026985 \pi)^{3}}} \nonumber\\
                  &   &= 3.01 \nonumber\label{eq:gc}
\end{eqnarray}


\end{sloppy}

\end{document}